# Electrocatalytic Performance of 2D Monolayer WSeTe Janus Transition Metal Dichalcogenide for Highly Efficient H$_2$ Evolution Reaction

**Vikash Kumar,[1] Shrish Nath Upadhyay,[2] Dikeshwar Halba[1] and Srimanta Pakhira[1,3*]**

[1] *Theoretical Condensed Matter Physics and Advanced Computational Materials Science Laboratory,* Department of Physics, Indian Institute of Technology Indore, Khandwa Road, Simrol, Indore-453552, MP, India.

[2] *Theoretical Condensed Matter Physics and Advanced Computational Materials Science Laboratory,* Department of Metallurgical and Materials Science (MEMS), Indian Institute of Technology Indore, Khandwa Road, Simrol, Indore-453552, MP, India.

[3] *Theoretical Condensed Matter Physics and Advanced Computational Materials Science Laboratory,* Centre for Advanced Electronics (CAE), Indian Institute of Technology Indore, Khandwa Road, Simrol, Indore-453552, MP, India.

*Corresponding author: spakhira@iiti.ac.in (or) spakhirafsu@gmail.com `

**ABSTRACT**

Now-a-days, the development of clean and green energy sources is the prior interest of research due to increasing global energy demand and extensive usage of fossil fuels that create pollutants. Hydrogen has the highest energy density by weight among all chemical fuels. For the commercial-scale production of hydrogen, water electrolysis is the best method which in turn requires an efficient, cost-effective and earth-abundant electrocatalyst. Recent studies have shown that the 2D Janus TMDs are highly effective in the electrocatalytic activity for HER. Herein we report a 2D monolayer WSeTe Janus TMD electrocatalyst for HER. We studied the electronic properties of 2D monolayer WSeTe Janus TMD using periodic DFT calculations, and the direct electronic band gap was obtained to be 2.39 eV. After the calculations of electronic properties, we explored the HER intermediates including various transition state structures (Volmer TS, Heyrovsky TS, and Tafel TS) using a molecular cluster model of WSeTe noted as $W_{10}Se_9Te_{12}$. The present calculations revealed that the 2D monolayer WSeTe Janus TMD is a potential electrocatalyst for HER. It has the lowest energy barriers for all the TSs among other TMDs, such as $MoS_2$, $Mn-MoS_2$, MoSSe, etc. The calculated Heyrovsky energy barrier (= 8.72 kcal.mol$^{-1}$) for the Volmer-Heyrovsky mechanism is larger than the Tafel energy barrier (=3.27 kcal.mol$^{-1}$) in the Volmer-Tafel mechanism. Hence our present study suggests that the formation of $H_2$ is energetically more favorable via the Vomer-Tafel mechanism. This work helps shed light on the rational design of effective HER catalysts.

**INTRODUCTION**

The energy crisis has become a major global concern in present days due to the rapid growth of the population, which increased the consumption of fossil fuels and their detrimental impact on the environment and has created an urgent need for a solution to meet the growing demand for clean energy. The quick depletion of fossil fuels has driven researchers to explore green and renewable energy alternatives to fossil fuels to tackle the problem of energy and environmental issues. Hydrogen has the advantages of comprehensive sources, high calorific value, non-polluting combustion, and recyclable regeneration, and it is also considered to be an excellent substitute for fossil fuels.[1] As a clean energy carrier in fuel cells, hydrogen is industrially produced from carbon feedstock through various processes. It is not environmentally friendly due to the release of $CO_2$ at the time of production of $H_2$. There are many ways to produce hydrogen, but electrolysis of water-splitting chemical reactions for hydrogen production is considered to be a reliable method because of its high purity, high efficiency, and no pollution in the whole process.[2,3] Hydrogen production from water electrolysis usually requires a catalyst to accelerate water splitting chemical reaction, in which currently many precious metals such as Pt and its alloy are being used. However, the large-scale application of hydrogen production by electrolysis of water is limited due to the expensive and scarce precious metal electrocatalysts.[4] Electrocatalysts are at the heart of the process because they are necessary for enhancing the reaction rates for producing $H_2$ and $O_2$ at the cathode and anode, respectively, via the hydrogen evolution reaction (HER) and oxygen evolution reaction (OER). An efficient catalyst can improve hydrogen production efficiency by reducing overpotential and accelerating the rate of HER.[5] Therefore, it is necessary to find alternative non-precious metal-based electrocatalysts to produce the $H_2$ at a commercial scale.

Two-dimensional (2D) materials have attracted research interest in recent years due to their excellent optical, electronic, and electrocatalytic properties and applications in various fields. Among the different types of 2D materials, transition metal dichalcogenides (TMDs) sheets, such as $MoS_2$, $WS_2$, $MoSe_2$, $WSe_2$ etc., have attracted much attention due to their excellent semiconducting, physical, and chemical properties.[6,7] Transition metal dichalcogenide (TMDs) (2D monolayers) have been fabricated under precise control by chemical vapor deposition (CVD), and they are considered promising alternatives to noble metal catalysts for effective HER due to their unique crystal structure and electronic properties.[6] However, TMDs materials are still less efficient for HER than Pt due to the limited density of active sites. Although large-scale studies have been carried out to increase the density of active sites, it is still beneficial and necessary to develop new strategies to enhance the catalytic activity of TMDs-based materials.[6,8,9] To overcome the limitations of pristine TMDs and to obtain excellent HER activity, new techniques are required to potentially improve the electrochemical performance of the HER by tuning the electrocatalytic properties of the pristine TMDs. Due to several constraints of TMDs, their overall electrocatalytic performance is limited, and the pristine TMDs have inert basal planes. The key challenge is to activate the basal plane of these TMDs by phase engineering, which includes doping of foreign atoms and creating defects. Recently, Ekka et al. computationally designed the Mn-doped $MoS_2$ (Mn-$MoS_2$) material to study the HER activity.[2] They developed a molecular cluster model system of the 2D single layer Mn-$MoS_2$ TMD noted as $Mn_1Mo_9S_{21}$. They found that the doping of the Mn atom in a 3x3 supercell of the 2D monolayer $MoS_2$ has reduced the band gap from 2.6 eV to 0 eV, and enough electron density appeared at the Fermi energy level. They reported that doping of the Mn atom enhanced the catalytic activity of $MoS_2$ towards HER.[2]

Recently another class of 2D materials that has recently gained much interest is the Janus materials. A few theoretical and experimental studies have been performed on 2D monolayer Janus TMDs. These materials have a structural configuration MXY, where M is a transition metal, and X and Y are different types of chalcogen elements (M = Mo, W *etc*., and X/Y = S, Se, and Te; X≠Y). The Janus TMDs show a lack of mirror symmetry and have vertical dipoles. The 2D Janus TMD (denoted as MXY) possesses a symmetry-broken structure with a $C_{3v}$ point group compared with the single-layer $MX_2$ TMD structure, which has a higher symmetry of the $D_{3h}$ point group.[10,11] An electronegativity difference between the X and Y layers of the Janus TMDs contributes to an inherent dipole moment. This property subtly distributes the holes and electrons present in its structure on its surface to enhance the electrocatalytic activity at room temperature.[12–15] Theoretically, the stability of group-VI chalcogenides MXY monolayers such as MoSSe, WSSe, WSeTe, and WSTe 2D monolayers has been checked and determined through phonon dispersion calculation and molecular dynamics simulations.[16,17] Recently, 2D Monolayer Janus MoSSe and WSSe materials have been broadly studied, which have proved very useful for their applications in sensors, electrochemistry, catalysis, and other devices.[1,18] Pakhira et al. theoretically studied the electrocatalytic activity of the 2D monolayer MoSSe Janus TMD towards HER, and they reported that it has the lowest energy barriers among several TMDs such as $MoS_2$, $WS_2$, and $W_xMo_{1-x}S_2$.[1]

Driven by this, in the present work, we systematically investigate the stability and hydrogen evolution activity of the 2D monolayer Janus WSeTe TMD by using the first principle-based Density Functional Theory (DFT) calculations. In this work, we employ first principles-based periodic density functional theory (DFT) with van der Waals (vdW) corrections to compute equilibrium geometry, structure, band structure, electronic band gap, Fermi energy level, and total density of state (DOS) of the 2D single-layer WSeTe Janus TMD.[19,20] Electronic properties calculations of the 2D monolayer Janus WSeTe show that this material may be excellent for $H_2$ production with high catalytic performance. Exposed Se-/Te-edge ($\bar{1}010$) and W-edge ($10\bar{1}0$) edges of the 2D monolayer Janus TMD WSeTe are found to be catalytically active for HER and while the Se-Mo-Te tri-layer of the Janus WSeTe TMD was the uncovered surface. To examine the reaction pathway of the HER mechanism, we have computationally designed a non-periodic finite molecular cluster model system (noted as $W_{10}Se_9Te_{12}$) of the 2D Janus WSeTe to study and explore the HER mechanism with the reaction pathway, as shown in Figure 1. The finite molecular cluster $W_{10}Se_9Te_{12}$ describes the W and Se/Te edges of the 2D monolayer Janus WSeTe, which are important for studying the HER mechanism. It has been shown that the DFT methods are very important for studying HER activity for $H_2$ production by electrochemical water splitting. They are very useful for obtaining the electronic properties, thermochemistry, reaction kinetics, reaction barriers, and activation energy constraints of the individual active sites of TMDs.[6,13,21–24] One of the main characteristics of calculating the flow of a reaction is the free energy **(ΔG)** changes of possible reaction intermediates during the reactions. Therefore, it is important to calculate the ΔG during hydrogen adsorption to screen suitable candidates among the available options, which is an important parameter for evaluating catalytic activity during the HER process. We computationally found that the Janus 2D WSeTe has exceptional electrochemical performance for the HER process and reasonable electrochemical parameters such as the low value of the Tafel slope, low activation energy barrier, and a large value of turnover frequency (TOF).

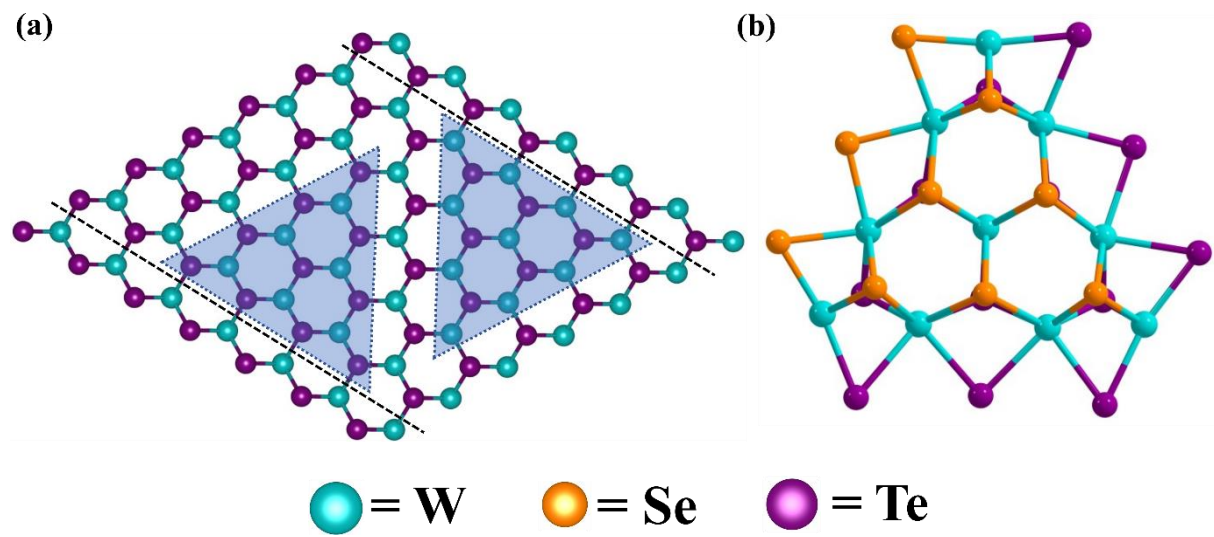

**Figure 1.** (a) Schematic presentation of the periodic 2D monolayer Janus WSeTe material. Two black dotted lines show the ($\bar{1}010$) Se-/Te-edges and ($10\bar{1}0$) W-edges of the 2D monolayer Janus WSeTe. Two triangles show the W-edges and Se-/Te-edges cluster model of the 2D monolayer WSeTe JTMD. (b) The W-edge finite molecular cluster model system ($W_{10}Se_9Te_{12}$) of the 2D WSeTe JTMD is presented here for studying the HER.

**METHODS AND COMPUTATIONAL DETAILS**

**(a) Periodic DFT calculations for 2D monolayer Janus WSeTe**

We have performed periodic DFT calculations to study the 2D monolayer structure and electronic properties of the 2D monolayer WSeTe JTMD. The electronic properties were studied at the equilibrium geometry of the 2D monolayer WSeTe JTMD. Therefore, we first performed geometry optimization to get the equilibrium lattice constants by using the B3LYP-D3 method.[1,2,24–27] At the equilibrium geometry of the WSeTe JTMD, the electronic properties calculations were carried out by applying the same B3LYP-D3 DFT-D method.[1,2,24–27] A*b initio*-based CRYSTAL17[28–30] suit code, which uses Gaussian type orbitals (GTOs), was used to obtain the equilibrium structure and electronic properties of the WSeTe. The earlier reports in the literature show that the localized GTO kind of basis sets are more accurate in solving the Hartree–Fock (HF) part of the Schrödinger equations implemented in the hybrid density functional methods.[22,24,31–34] Grimme's semi-empirical dispersion corrections (-D3) were incorporated in the periodic B3LYP DFT method to consider the weak van der Waals (vdW) interactions between the layers and atoms of the 2D monolayer WSeTe JTMD material.[35–38]

Triple-ζ valence with polarization (TZVP) quality Gaussian types of atomic basis sets were applied for the W, Se, and Te atoms in the present calculations.[39–41] The DFT-D methods provide a good quality of geometry of the 2D layered structure materials by reducing the spin contamination effects, and it does not affect the calculations of electronic structure and property calculations of the 2D WSeTe JTMD.[42–45] Spin-polarized calculations have been performed to obtain the equilibrium structures and to study the electronic properties during periodic hybrid DFT-D calculations. A spin-polarized solution has been computed after the definition of the (α, up spin and β, down spin) electron occupancy. In other words, it may be here noted that spin-unrestricted wave functions are used in the present calculations to incorporate spin polarization by using the keywords SPIN and SPINLOCK in *ab initio* CRYSTAL17 program. The ATOMSPIN keyword is also used to specify the single spin of the TM atoms in the material.[32,46] A threshold value of $10^{-7}$ a.u. was used for the convergence of forces, energy, and electron density for all cases. The periodicity in the z-direction of the crystal structure was ignored by keeping the height of the unit cell around 500 Å, i.e., the vacuum region of approximately 500 Å was considered in the present calculations to accommodate the vacuum environment. The vacuum region of 500 Å was set to avoid the interlayer interaction between two consecutive layers in the z-direction.[28]

Electronic structure calculations were performed on the optimized geometry of the 2D monolayer WSeTe JTMD material by employing the same level of theory. Monkhorst k-mesh grids of the size 15x15x1 were used to compute the 2D electronic layer structure, geometry, band structure, electronic band gap ($E_g$), Fermi energy level, and total density of states (DOS).[47] In the first Brillouin zone, a total eight number of electronic bands were calculated around the Fermi energy level in a high symmetric ***Γ-M-K-Γ*** direction. All the atomic orbitals of the W, Se, and Te atoms were taken into account to calculate the total density of states of the 2D monolayer WSeTe JTMD. The electrostatic potential effects were added in the present calculations, i.e., the electronic band structures and DOS were calculated with respect to the vacuum, and band alignment has been conducted. VESTA,[48] a visualization software was used to create graphics and visualize the equilibrium 2D monolayer structure of the 2D monolayer WSeTe JTMD.

**(b) Finite Cluster Modelling**

In the present study, we have designed a finite non-periodic molecular cluster model (noted by $W_{10}Se_9Te_{12}$) of the 2D monolayer WSeTe JTMD to investigate the HER pathway and explore the chemical reaction mechanism. A schematic representation of the finite molecular cluster model system $W_{10}Se_9Te_{12}$ is shown in Figure 1b. We have developed a molecular cluster model for the 2D WSeTe JTMD, following a similar approach as the previously published design of the 2D monolayer Janus WSSe TMD.[18] Our molecular cluster model allows us to investigate the unique properties and potential applications of the WSeTe JTMD structure. The development of the finite molecular cluster model was done in such a way that it has the same chemical properties as the periodic 2D slab of WSeTe JTMD. We have used the M06-L DFT method to determine the reaction barriers, kinetics, and bond energies. This method provides more flexibility in accuracy while using molecular cluster and gives very precise results for the HER calculations.[49,50] It is easy to use a cluster model with net charges, which is not possible to consider in the case of a periodic system using the CRYSTAL17 suite code. Incorporation of simultaneous electrons ($e^-$) and protons ($H^+$) while performing the reaction studies of various steps in HER becomes very easy when the molecular cluster model is used. All the non-periodic calculations were performed by using the GAUSSIAN16[51,52] suit code. We focused on the energy barriers and changes in free energy during the reaction to explore the reaction pathways by employing Minnesota Density Functional based on the meta-GGA approximation, which is intended to be good and fast for transition metal complexes. The LANL2DZ[53,54] Gaussian type of atomic basis sets with the relativistic effective core potentials (ECPs) have been used for the W, Se, and Te atoms and the 6-31+G** (double-ζ Pople-type)[55,56] Gaussian basis sets for the H and O atoms in the present computations. TS structures (transition state structures or saddle point) of $H^*$-migration, Heyrovsky, and Tafel reaction steps were obtained to study the reaction barriers, which was confirmed by obtaining one imaginary frequency, modes of vibration, and intrinsic reaction coordinate (IRC) calculations.[52,57–59] ChemCraft software was used to visualize them.[60]

Harmonic vibrational and thermodynamic calculations were performed to get the relative Gibbs free energy (**ΔG**) and enthalpy (**ΔH**). All the reaction intermediates involved in the subject reaction (HER) were confirmed to be stable and feasible by obtaining all the positive frequencies (except in the case of TSs, where one imaginary frequency was obtained) during the frequency calculations. PCM (polarization continuum model) calculations were performed to study the solvation effects on reaction barriers. For the PCM calculations, water was taken as a solvent with the dielectric constant of 80.13.[61] PCM is one of the best models to consider the solvation effects, and it is a commonly used method in computational quantum chemistry to model solvation effects.[6,34,62]

**(c) Theoretical calculations and equations**

The electrocatalytic performances of the 2D monolayer WSeTe JTMD were characterized by computing the changes of the Gibbs free energy ($\Delta G$) for $H_2$ adsorption on the ($\bar{1}010$) S-/Se edges and ($10\bar{1}0$) W-edges of the system. The change of various energies, such as a change in Gibbs free energy ($\Delta G$), enthalpy ($\Delta H$), and electronic energy ($\Delta E$) were calculated at pH = 0, using the following equation

$$\Delta G = \sum G_{Products} - \sum G_{Reactants}$$

$$\Delta H = \sum H_{Products} - \sum H_{Reactants}$$

$$\Delta E = \sum E_{Products} - \sum E_{Reactants}$$

Where G, H, and E are Gibbs free energy, enthalpy, and electronic energy of the systems considered in the present study. The Gibbs's free energy of an electron ($e^-$) has been calculated at the standard hydrogen electrode (SHE) conditions where electron ($e^-$) and proton ($H^+$) (pH = 0) are in equilibrium with 1 atm $H_2$. The PCM analysis was employed for all the theoretical calculations to describe the solvation effects by employing the M06-L DFT method, and three $H_2O$ molecules with one $H_3O^+$ (i.e., $3H_2O\_H_3O^+$) have been combined unambiguously for the Heyrovsky reaction step.

## RESULTS AND DISCUSSION

## Structural and Electronic Properties Calculations of the 2D monolayer WSeTe JTMD.

The catalytic performance depends on the stability of the material and its electronic properties of it, so for the stability of the material, we first look at the structural and electronic properties of the 2D monolayer WSeTe JTMD. We computationally designed a 2D monolayer structure of the Janus WSeTe TMD, which is a highly symmetric slab with (001) basal plane, and the symmetrical unit structure of the 2D monolayer slab has been used for the structural and electronic properties calculations. It was computationally found that the Janus WSeTe monolayer slab has a *P$\bar{3}$m1* symmetry with symmetric Int layer group number 69 corresponding to the tetragonal 2D layer system. The top view and side view of equilibrium 2D single structures of the Janus WSeTe TMD are shown in Figure 2. The Janus WSeTe monolayer slab structure loses out-of-plane symmetry compared to the 2D pristine monolayer $MoS_2$, $WS_2$, and $WSe_2$ TMDs.[6,58] 2D Janus WSeTe monolayer can be created by replacing one layer of Se atoms with Te atoms in the $WSe_2$ monolayer. We first consider the key properties of the 2D Janus WSeTe monolayer by examining its electronic equilibrium structure with its band structure, the position of the Fermi level, and the total density of states (DOS).

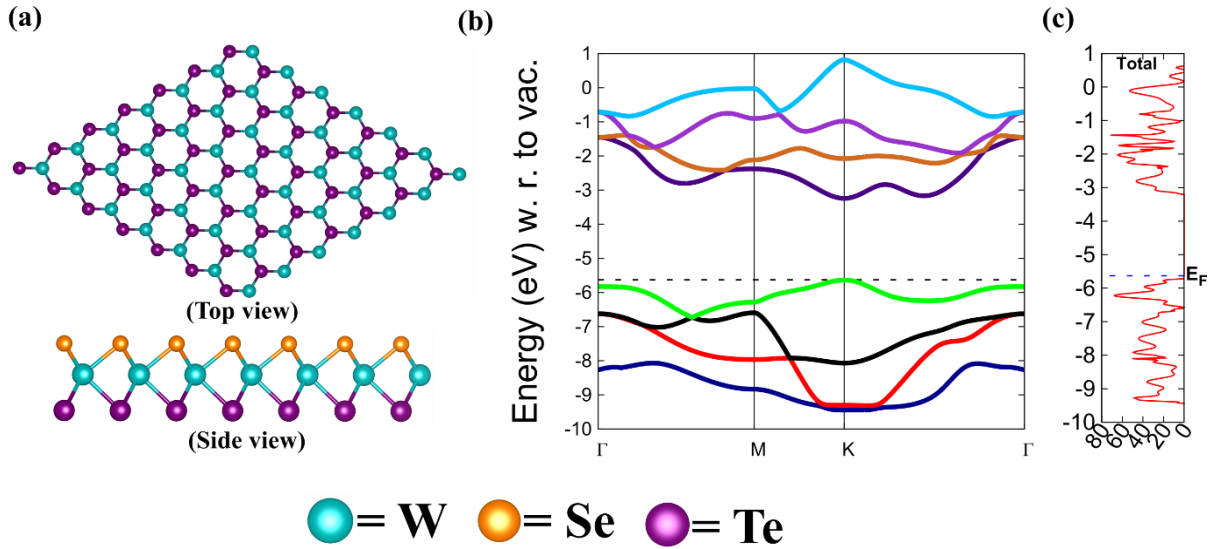

**Figure 2.** (a) Top and side view of the equilibrium structure of the 2D monolayer WSeTe Janus TMD. (b) Band structure and (c) the total density of states (DOS) of the 2D monolayer WSeTe Janus TMD.

In contrast to ordinary TMDs ($WSe_2$ and $WTe_2$ monolayers), the dissimilarity in the electronegativity between Te (~2.10) and Se (~2.55) atoms in the WSeTe monolayer structure of the Janus TMD results in unequal bond lengths of the W-Se and W-Te bonds, as shown in Table 1.[63] It was evident that the W-Te bond length is still 0.17 Å longer than the W-Se bond. The bond angle between the W-Se and W-Te is about 83.29°. Equilibrium geometry, band structure, and total density of states (DOS) of the 2D monolayer WSeTe JTMD were computed by the B3LYP-D3 DFT-D method, as displayed in Figure 2. The lattice parameters are consistent with the previous report[64] and the present study found that the lattice parameters of the JTMD are a = b = 3.37 Å obtained by the same DFT-D method. The electronic band structure of the 2D WSeTe is shown in Figure 2b, which indicates that it has semiconducting properties with a direct band gap about ($E_g$) 2.39 eV at K point, which is in good agreement with the earlier reported data.[13,63–66] The minimum conduction band (CB) and the maximum valence band (VB) are directly located at the *K* point, Therefore, it is a direct band gap semiconductor in nature. To obtain the fine band structure of the 2D monolayer Janus WSeTe material, a total number of eight electronic energy bands were plotted in specific directions of the irreducible Brillouin zone by selecting the ***Γ-M-K-Γ*** high symmetry path, as depicted in Figure 2b, which is consistent with the 2D layer group symmetry of the Janus TMD materials. The Fermi level ($E_F$) is at -5.63 eV, which is very close to the top of the valence band (VB), and the direct electronic band gap is located at the K point in the band structure calculation, which was later confirmed by the total DOS calculations. Figure 2c represents the total density of states of the pristine 2D monolayer Janus WSeTe. The maximum value of the VB and the minimum value of the CB are -5.63 eV and -3.24 eV, respectively, in the band structure diagram (i.e., E-k diagram) depicted in Figure 2b. From the total DOS, we observed that there is an electron density of states around the Fermi level ($E_F$) with a direct band gap about 2.39 eV at K point, which makes 2D monolayer Janus WSeTe semiconducting in nature, and the JTMD may help to enhance the electrocatalytic activity for HER. So, this material 2D monolayer Janus WSeTe TMD can be expected as a catalyst towards HER.

**Table 4.** The calculated lattice parameters, bond angle, bond distance, and band gap of the 2D monolayer Janus WSeTe TMD at the equilibrium position.

| Materials | Lattice parameter (Å) | ∠SeWTe (in °) | Bond distance (in Å) | | Band Gap (eV) | References |
|---|---|---|---|---|---|---|
| | | | W-Se (Å) | W-Te (Å) | | |
| WSeTe | $a = b = 3.37$ | 83.29 | 2.52 | 2.69 | 2.39 | This work |
| WSeTe (Previous Report) | $a = b = 3.45$ | 81.68 | 2.56 | 2.72 | 1.79 | 13,63 |

**HER MECHANISMS**

To study the catalytic activity of the 2D monolayer WSeTe Janus TMD towards the HER, the relative free energy of hydrogen adsorption ($\Delta G_H$) was calculated by employing the DFT method. For better catalytic activity, $\Delta G_H$ should be almost equal to 0 eV, indicating that the binding of hydrogen to WSeTe JTMD is neither too strong nor too weak, i.e., it should be moderate.[2,67,68] The entire reaction pathway of HER follows either the Volmer-Heyrovsky or Volmer-Tafel mechanism independent of the medium. These mechanisms are given in the following equations.

$$\text{Volmer Reaction} \quad\quad M^* + H^+ + e^- \rightarrow H^*_{ads} \quad\quad (i)$$
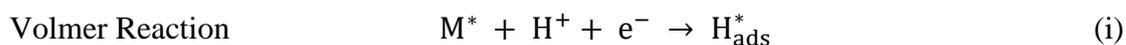

Heyrovsky Reaction $\quad\quad\quad\quad\quad\quad H_{ads}^* + H^+ + e^- \rightarrow H_2 + M^* \quad\quad\quad$ (ii)

Tafel Reaction $\quad\quad\quad\quad\quad\quad\quad 2H_{ads}^* \rightarrow H_2 + M^* \quad\quad\quad\quad\quad\quad$ (iii)

The first step is the Volmer reaction, which involves the adsorption of H* by transferring an electron and proton at the surface-active sites, as shown in equation (i). In the next step, the Heyrovsky reaction involves a process on the electrode surface in which adsorbed hydrogen atoms ($H_{ads}^*$) combine with hydronium ions in an acid solution and the formation of the $H_2$ occurs, as shown in equation (ii). One other step of $H_2$ formation is called the Tafel reaction step, in which two adsorbed hydrogen atoms ($H_{ads}^*$) combine to form the $H_2$, as equation (iii) shows.

To investigate the HER performance of 2D monolayer WSeTe JTMD, we developed a finite non-periodic cluster model system $W_{10}Se_9Te_{12}$ and explored the reaction pathways, mechanisms, thermodynamics, chemical kinetics, transition state (TS) structures, intermediates, and reaction barriers. Here, we studied both the Volmer-Heyrovsky and Volmer-Tafel HER mechanisms, as shown in Figures 3 and 4 of the HER pathways. We followed both Figures 3 and 4 for the $H_2$ formation and calculated the different intermediates and transition states (TSs) during the HER process. Since the complete HER process may proceed through either the Volmer-Heyrovsky reaction mechanism or the Volmer-Tafel reaction mechanism, it is necessary to examine the energy barriers of individual reaction steps to trace the rate-limiting step. Since the electrocatalytic properties of 2D monolayer Janus WSeTe materials have been less studied, we studied the reaction pathway proposed for HER following by Vollmer-Heyrovsky and Vollmer-Tafel to predict the most prominent mechanism of 2D monolayer Janus WSeTe TMD materials.

**Figure 3.** Volmer-Heyrovsky hydrogen evolution reaction (HER) mechanism pathway on the 2D monolayer Janus WSeTe surfaces.

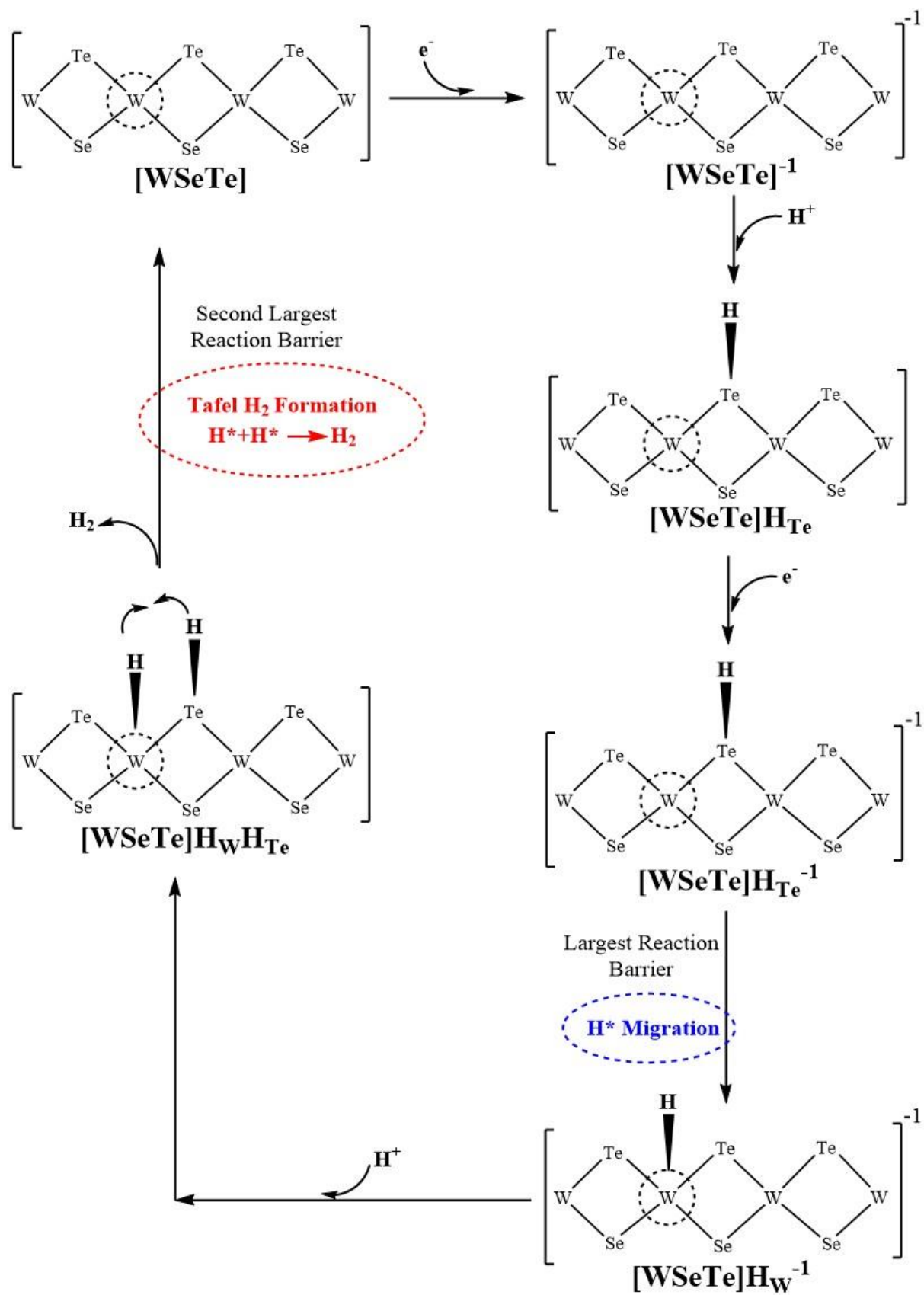

**Figure 4.** Volmer-Tafel hydrogen evolution reaction (HER) mechanism pathway on the 2D monolayer Janus WSeTe surfaces.

**VOLMER-HEYROVSKY MECHANISM**

The Volmer-Heyrovsky mechanism goes along with a two-electron transfer process, and the whole reaction pathway of this mechanism (when the HER occurs at the W edge of the 2D monolayer WSeTe non-periodic molecular cluster model) is shown in Figure 3. This multi-step electrode reaction schematic diagram includes possible intermediates and transition states (TSs) formation in the proposed reaction mechanism throughout the HER process. First, we have calculated the free energies of the most probable intermediates as a basis for describing thermodynamics, i.e., the kinetics of the HER. Then, we examined barriers at individual reaction steps to locate rate-limiting steps. The first reaction step is the Volmer reaction step, where protons ($H^+$) interact with the electrons, and the hydrogen atoms get adsorbed on the active region of the catalyst (in short, $* + H^+ + e^- \rightarrow H^*_{ads}$, where * represents the active sites of the catalyst).

Similarly, in the Heyrovsky reaction step, the formation of $H_2$ plays a crucial role through the participation of $H^*$ at the transition metal site and the requirement of a proton from the adjacent hydronium ion ($H_3O^+$). Using the finite molecular cluster model $W_{10}Se_9Te_{12}$, we have combined the number of electrons and protons independently in discrete steps. The Volmer-Heyrovsky reaction mechanism of the proposed HER on the active edge of the 2D monolayer Janus WSeTe TMD has been thoroughly investigated, and various successive reaction steps with Gibbs free energy changes have been calculated on the same level of the DFT-D method. The relative free energy of each intermediate reaction step has been estimated and listed in Table 2. The detailed reaction steps involved in this proposed HER pathway are as follows.

I. To initiate the HER process, an electron is absorbed into the $W_{10}Se_9Te_{12}$ [WSeTe] cluster model, resulting in a negatively charged cluster [WSeTe]$^{-1}$ solvated in water with a delocalized electron on its surface. Here, we found that the first reduction potential for introducing a single electron calculated by the DFT method is about -509.09 mV. The free energy of the electron $G(e^-)$ is calculated by the expression $G(e^-)= G (1/2\ H_2)-G(H^+)$. The free energy $G(H^+)$ of the proton is -274.86 kcal.mol$^{-1}$, taken from the value previously reported by Tissandier et al.[69]. The equilibrium geometries of the [WSeTe] and [WSeTe]$^{-1}$ are shown in Figures 5a and 5b, respectively. The free energy change between [WSeTe]$^{-1}$ and [WSeTe] is found to be 11.74 kcal.mol$^{-1}$, which is listed in Table 2.

II. The next step involves the addition of an $H^+$ to the most active Te edge of 2D $[WSeTe]^{-1}$, which results in the formation of the next reaction intermediate, $[WSeTe]H_{Te}$. The equilibrium geometry of the $[WSeTe]H_{Te}$ is shown in Figure 5c. The free energy cost ΔG of the first protonation of $[WSeTe]^{-1}$ to obtain the $[WSeTe]H_{Te}$ intermediate is 2.53 kcal.mol$^{-1}$, as shown in Table 2.

III. Further, an electron is added to $[WSeTe]H_{Te}$ resulting in a second reduction of $[WSeTe]H_{Te}^{-1}$, and the reduction potential of this step is calculated to be approximately -796.17 mV. The energy cost of the $[WSeTe]H_{Te}$ to obtain the $[WSeTe]H_{Te}^{-1}$ intermediate is ΔG = 18.67 kcal.mol$^{-1}$, as shown in Table 2. After adding another electron, the equilibrium geometry of the $[WSeTe]H_{Te}^{-1}$, is shown in Figure 5d.

IV. In the next step, the hydride ion ($H^*$) adsorbed on the Te atom site migrates to the adjacent W atom site to form the transition state of $[WSeTe]H_W^{-1}$, known as the $H^*$-migration reaction step (TS1). Harmonic imaginary vibrational frequency analysis and internal feedback coordination (IRC) calculations have been performed to confirm TS1. In the TS1 calculation, where $H^*$ migrates from the Te atom site to the adjacent W atom site, we found an imaginary frequency around -494.71 cm$^{-1}$. The equilibrium geometry of the $H^*$-migration reaction step (TS1) is shown in Figure 5e. The energy cost to obtain the $H^*$ migration from the Te atom site to the W atom site in the TS1 intermediate is ΔG = 3.80 kcal.mol$^{-1}$, as shown in Table 2.

V. In the next step after the calculation of TS1, the energy cost to obtain $[WSeTe]H_W^{-1}$ from TS1 is ΔG = -24.01 kcal.mol$^{-1}$, as shown in Table 2. The equilibrium geometry of this $[WSeTe]H_W^{-1}$ is shown in Figure 5f.

VI. Further, the HER proceeds by the addition of a second $H^+$ from the medium to the Te site of $[WSeTe]H_W^{-1}$, resulting in the formation of the next intermediate, $[WSeTe]H_W H_{Te}$, with an energy cost (ΔG) of -0.62 kcal.mol$^{-1}$. The equilibrium geometry of the $[WSeTe]H_W H_{Te}$ is shown in Figure 5g.

VII. The next step in the HER mechanism is the formation of $H_2$ by the Heyrovsky mechanism, in which a hydrated hydrogen water cluster ($3H_2O+H_3O^+$) is added to [WSeTe]$H_WH_{Te}$. The energy cost of this step to obtain [WSeTe]$H_WH_{Te}$+3$H_2O$+$H_3O^+$ from [WSeTe]$H_WH_{Te}$ is $\Delta G$ = -3.88 kcal.mol$^{-1}$, as shown in Table 2. After adding the water cluster, the equilibrium geometry of the [WSeTe]$H_WH_{Te}$+3$H_2O$+$H_3O^+$ is shown in Figure 5h.

VIII. A second transition state, called Heyrovsky's transition state (TS2), as shown in Figure 5i, was studied. The second TS arises from [WSeTe]$H_WH_{Te}$+3$H_2O$+$H_3O^+$, where H$^*$ from the W site and H$^+$ from the water group combine and, the formation and evolution of $H_2$ occur from the surface of the electrocatalyst. The activation energy barrier of Heyrovsky TS2 is calculated to be approximately 5.95 kcal.mol$^{-1}$ in the gas phase, as shown in Table 2. During the TS2, we found an imaginary frequency around -839.81 cm$^{-1}$, confirming this structure to be a transition state structure.

IX. After the formation of Heyrovsky TS (TS2), the system becomes [WSeTe]$H_{Te}^{+1}$. An $H_2$ molecule and 4$H_2O$ molecules evolved, and the free energy consumption is 9.94 kcal.mol$^{-1}$, as shown in Table 2. After forming the $H_2$ molecule equilibrium geometry of the [WSeTe]$H_{Te}^{+1}$, is shown in Figure 5j.

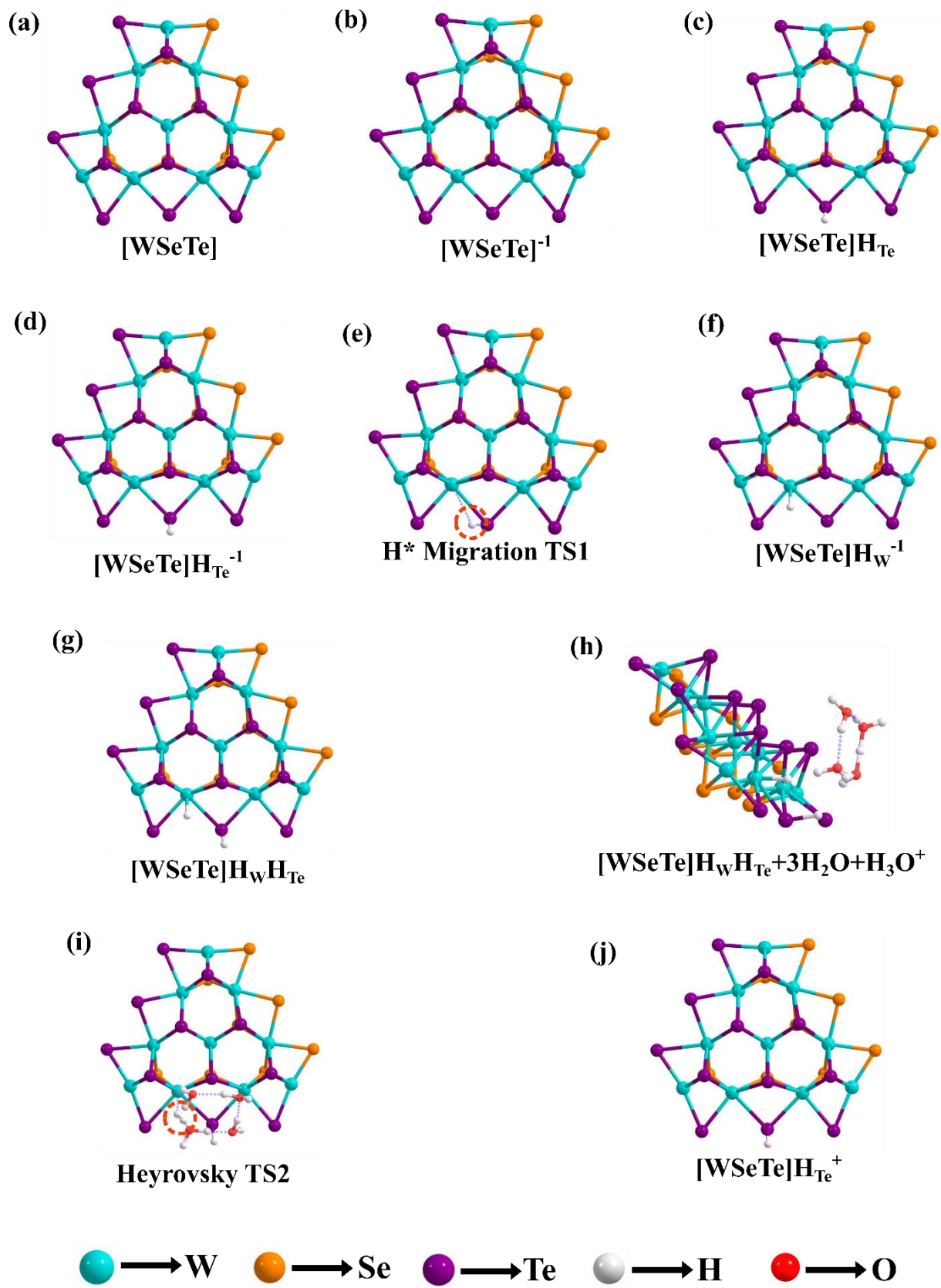

**Figure 5.** Equilibrium geometries of 2D Janus WSeTe during the Volmer-Heyrovsky mechanism for the HER computed by the M06-L DFT method are shown here. HER happens at the Te-terminated W-edge of the 2D monolayer WSeTe JTMD surface.

In conclusion, we have calculated two transition structures, TS1 and TS2 (Volmer TS and Heyrovsky TS), the first of which is when H* migrates from the chalcogen (Te) site to a metal site (here, W). The second TS is calculated at the time of formation of the H$_2$ molecule on the active surface of the 2D monolayer Janus WSeTe material during the Heyrovsky reaction step in the HER process. The calculated Volmer and Heyrovsky reaction barriers in the gas-phase are 3.80 kcal.mol$^{-1}$ and 5.95 kcal.mol$^{-1}$, respectively. The changes in electronic energy (**ΔE**), enthalpy (**ΔH**), and free energy (**ΔG**) of the different reaction intermediates involved in HER (in gas phase calculations) are reported in Table 2.

**Table 2.** The energy changes (**ΔE, ΔH,** and **ΔG**) of various reaction intermediates and transition state structures (TSs) during the Volmer-Heyrovsky reaction mechanism computed by the M06-L DFT method in gas phase calculations are listed here.

| HER Reaction Intermediates | ΔE (kcal.mol$^{-1}$) Gas Phase | ΔH (kcal.mol$^{-1}$) Gas Phase | ΔG (kcal.mol$^{-1}$) Gas Phase |
|---|---|---|---|
| [WSeTe] → [WSeTe]$^-$ | 13.36 | 13.29 | 11.74 |
| [WSeTe]$^-$ → [WSeTe]H$_{Te}$ | -3.17 | 1.41 | 2.53 |
| [WSeTe]H$_{Te}$ → [WSeTe]H$_{Te}{}^-$ | 18.80 | 18.73 | 18.67 |
| [WSeTe]H$_{Te}{}^-$ → H*Migration TS1 | 4.21 | 3.30 | 3.80 |
| H*Migration TS1 → [WSSe]H$_W{}^-$ | -26.01 | -24.56 | -24.01 |
| [WSeTe]H$_W{}^-$ → [WSeTe]H$_W$H$_{Te}$ | -5.16 | -0.34 | -0.62 |
| [WSeTe]H$_W$H$_{Te}$ → [WSeTe]H$_W$H$_{Te}$ + 3H$_2$O + H$_3$O$^+$ | -16.27 | -16.59 | -3.88 |
| [WSeTe]H$_W$H$_{Te}$ + 3H$_2$O + H$_3$O$^+$ → Heyrovsky TS2 | 8.89 | 6.56 | 5.95 |
| Heyrovsky TS2 → [WSeTe]H$_{Te}{}^+$ | 6.60 | 6.92 | -9.64 |

The HER mechanism mainly focuses on two steps, according to Figure 6. One of them is the H* migration from the Te site to the W site, and the second one is hydrogen molecule formation through the water cluster. The potential energy surface (PES) of this Volmer-Heyrovsky reaction mechanism is shown in Figure 6. The change in Gibbs free energy (ΔG) relative to the progress of the reaction steps involved in the Volmer-Heyrovsky reaction has been plotted in Figure 6.

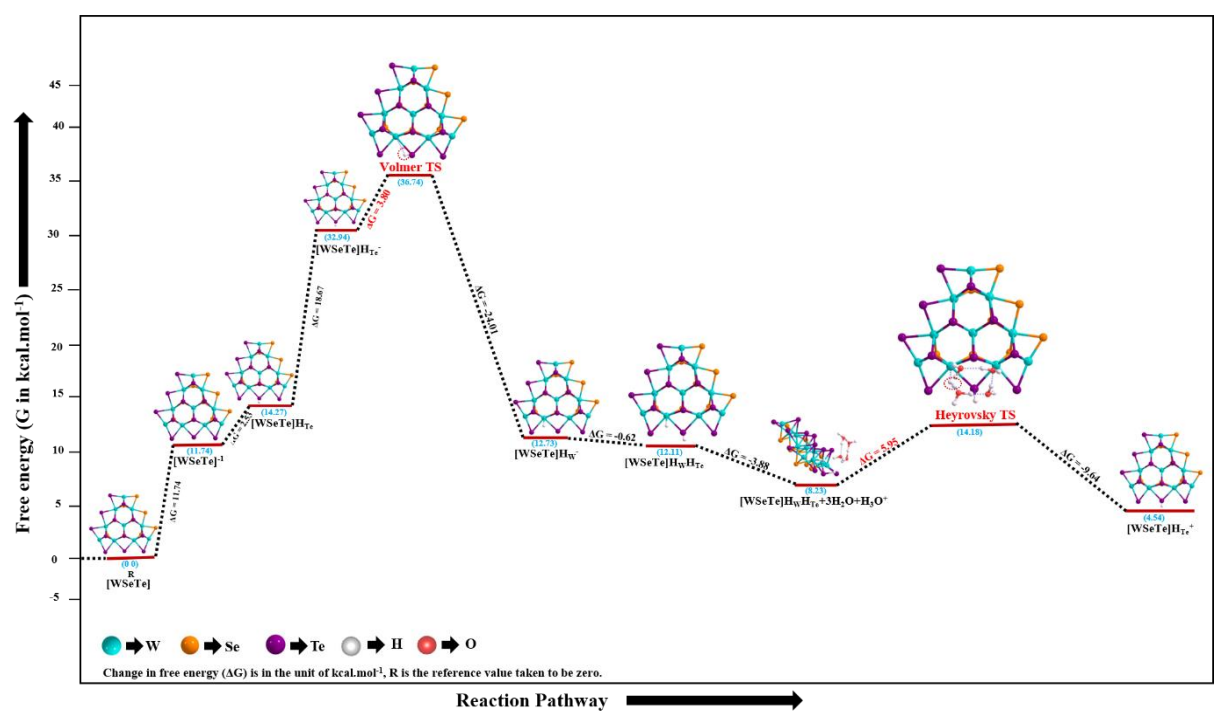

**Figure 6.** The PES of the Volmer-Heyrovsky reaction mechanism during HER on the surface of 2D monolayer Janus WSeTe material is presented here. Changes in free energy (ΔG) (in the gas phase) are expressed in kcal.mol$^{-1}$ in the plot.

As in the commercial field, most reactions are carried out in solution, so solvent-phase calculations to determine activation barriers are also required.[1,2,6] Therefore, in the present study, the solvent phase calculations are also performed by considering the solvent effect of the universal solvent "water" for the Volmer-Heyrovsky mechanism. We used the Polarizable Continuum Model (PCM) method to calculate the solvent phase energy barriers. This is the standard method for computational study to calculate the solvent effect. We used water as a solvent for the PCM calculations with a dielectric constant of 80.15. The present DFT-D calculations for the solvent phase were performed using the M06-L DFT method and found that the energy barrier for H*-migration (TS1) of the Volmer reaction step in the solvent phase is about 3.65 kcal.mol$^{-1}$. The lower energy barrier value of TS1 in the solvent phase calculation shows a better hydrogen adsorption on the surface of 2D monolayer WSeTe Janus TMD. Moreover, the energy barrier of TS2 in the Heyrovsky reaction step for H$_2$ formation is about 8.72 kcal.mol$^{-1}$ in the solvent phase. The changes in electronic energy, enthalpy, and Gibbs free energy of the Volmer and Heyrovsky reaction barriers in the solvent phase are given in Table 3. Here we observe that the 2D Janus WSeTe may be an excellent electrocatalyst for the HER.

**Table 3.** The energy changes ($\Delta E$, $\Delta H$, and $\Delta G$) of the transition states (TS) during the Volmer-Heyrovsky reaction mechanism by computed M06-L DFT method in solvent phase calculations are listed here.

| Activation barrier | $\Delta E$ (kcal.mol$^{-1}$) in solvent phase | $\Delta H$ (kcal.mol$^{-1}$) in solvent phase | $\Delta G$ (kcal.mol$^{-1}$) in solvent phase |
|---|---|---|---|
| **Volmer reaction barrier** | 4.07 | 3.16 | 3.65 |
| **Heyrovsky reaction barrier** | 11.93 | 9.33 | 8.72 |

**VOLMER-TAFEL REACTION MECHANISM**

H$_2$ formation can take place via two possible HER mechanisms; Volmer-Heyrovsky and Volmer-Tafel. The Volmer-Heyrovsky reaction mechanism has already been discussed. Now, we have also studied the Volmer-Tafel reaction mechanism of the proposed HER on the active edge of the 2D monolayer Janus WSeTe TMD. The Volmer-Tafel reaction mechanism has been thoroughly investigated, and various successive reaction steps with Gibbs free energy changes have been calculated on the same level as the DFT-D method. The Volmer-Tafel mechanism is also a two-electron transfer process for the H$_2$ formation. In the case of the Volmer-Tafel reaction mechanism, two adjacent hydrogens adsorbed on the catalyst surface recombine to form H$_2$, i.e., H$^*$+H$^*$→H$_2$, without further solvation of protons as in the case of Volmer-Heyrovsky mechanism. Figure 4 represents the complete reaction process involved in this proposed Volmer-Tafel reaction pathway at the surface of the 2D WSeTe JTMD. In the Vomer-Tafel mechanism, completion of the HER mechanism occurs in the following steps:

I. This Volmer-Tafel mechanism follows the same path as in the case of the Volmer-Heyrovsky mechanism until the [WSeTe]$H_W H_{Te}$ intermediate is generated from the [WSeTe]$H_W^{-1}$ with the gradual addition of a proton (H$^+$) at a free energy cost of -0.62 kcal.mol$^{-1}$, as Shown in the Volmer-Heyrovsky mechanism.

II. In the upcoming step, the two adsorbed hydrogens, one at the W atom site and the other at the Te atom site, both hydrogens recombine to form H$_2$, which is called the Volmer-Tafel mechanism. The activation energy barrier of the Tafel TS3 is calculated to be approximately 2.99 kcal.mol$^{-1}$ in the gas phase, as shown in Table 4. During the calculation of TS3, we found an imaginary frequency around -432.30 cm$^{-1}$, confirming this structure to be a transition state. The third TS arises from [WSeTe]$H_W H_{Te}$, where H$^*$ from the W atom site and H$^*$ from the Te atom site recombine, evolve into H$_2$ and dissociate from the system, as shown in Figure 7.

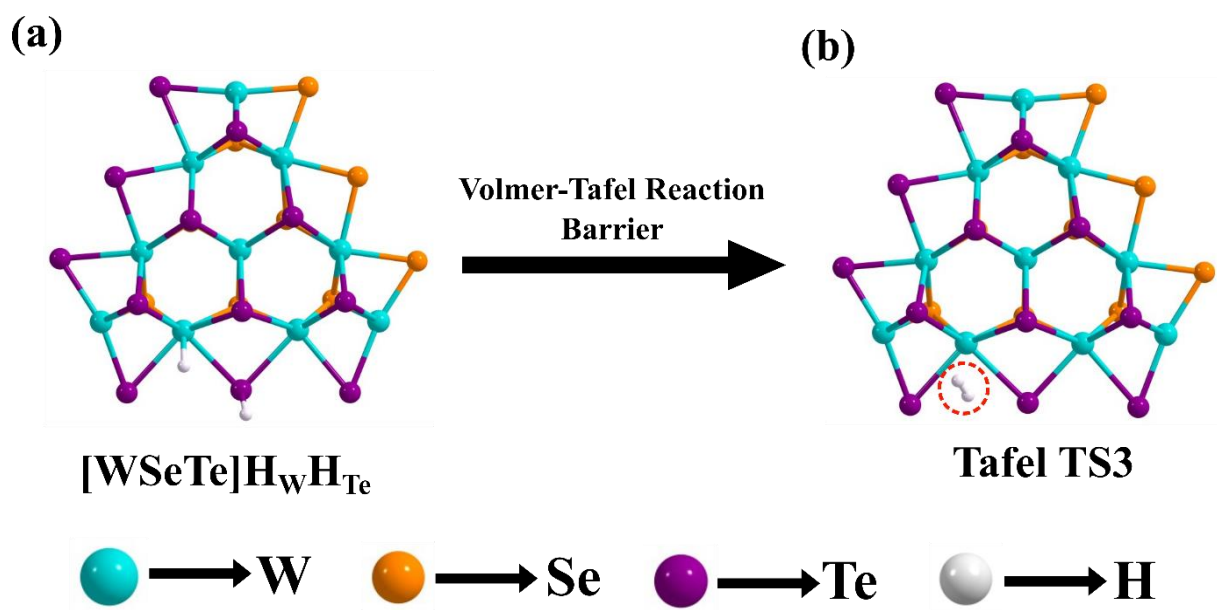

**Figure 7.** The Volmer-Tafel reaction mechanism equilibrium geometries of the [WSeTe]$H_W H_{Te}$ and TS3 are shown here.

III. After the formation of Tafel TS3, the initial step [WSeTe] of the mechanism is achieved. An H$_2$ molecule is evolved from the surface of the catalyst, and the free energy consumption is 15.16 kcal.mol$^{-1}$, as shown in Table 4.

In conclusion, we calculated two transition structures, TS1 and TS3 (H*-migration TS and Tafel TS), the first of which is when H* migrates from the tellurium (Te) site to a metal site (here, W). The third TS is calculated when the formation of an $H_2$ molecule occurs on the active surface of the 2D monolayer Janus WSeTe material during the Tafel reaction step in the HER process. The H*-migration and Tafel reaction barriers are 3.80 kcal.mol$^{-1}$ and 2.99 kcal.mol$^{-1}$, respectively, in the gas phase. The changes in electronic energy (**ΔE**), enthalpy (**ΔH**), and free energy (**ΔG**) of the different reaction intermediates involved in HER in gas phase calculations are reported in Table 4.

**Table 4.** The energy changes (**ΔE, ΔH,** and **ΔG**) of different intermediates and transition states (TS) during the Volmer-Tafel reaction mechanism computed by the M06-L DFT method in gas phase calculations are listed here.

| HER Reaction Intermediates | ΔE (kcal.mol$^{-1}$) Gas Phase | ΔH (kcal.mol$^{-1}$) Gas Phase | ΔG (kcal.mol$^{-1}$) Gas Phase |
|---|---|---|---|
| [WSeTe] → [WSeTe]$^-$ | 13.36 | 13.29 | 11.74 |
| [WSeTe]$^-$ → [WSeTe]H$_{Te}$ | -3.17 | 1.41 | 2.53 |
| [WSeTe]H$_{Te}$ → [WSeTe]H$_{Te}$$^-$ | 18.80 | 18.73 | 18.67 |
| [WSeTe]H$_{Te}$$^-$ → H*Migration (TS1) | 4.21 | 3.30 | 3.80 |
| H*Migration (TS1) → [WSSe]H$_W$$^-$ | -26.01 | -24.56 | -24.01 |
| [WSeTe]H$_W$$^-$ → [WSeTe]H$_W$H$_{Te}$ | -5.16 | -0.34 | -0.62 |
| [WSSe]H$_W$H$_{Te}$ → Tafel TS3 | 4.14 | 3.17 | 2.99 |
| Tafel TS3 → [WSeTe] | -5.32 | -5.77 | -15.16 |

We followed the Volmer-Tafel reaction pathway but mainly focused on two points, according to Figure 8. One of them was the migration of H* from the Te-site to the W-site, and the second was hydrogen formation through the two adsorbed hydrogen on the W-site and Te-site. The potential energy surface (PES) of this Volmer-Tafel reaction mechanism is shown in Figure 8. Figure 8 presents the change in Gibbs free energy (**ΔG**) relative to the progress of the reaction steps involved in the Volmer-Tafel reaction calculated in the gas phase. The solvation effect study was also performed during the HER by incorporating the polarization continuum model (PCM) into TS3. The free energy barrier of TS3 in the Tafel reaction step for hydrogen evolution is about 3.27 kcal.mol$^{-1}$ in the solvent phase.

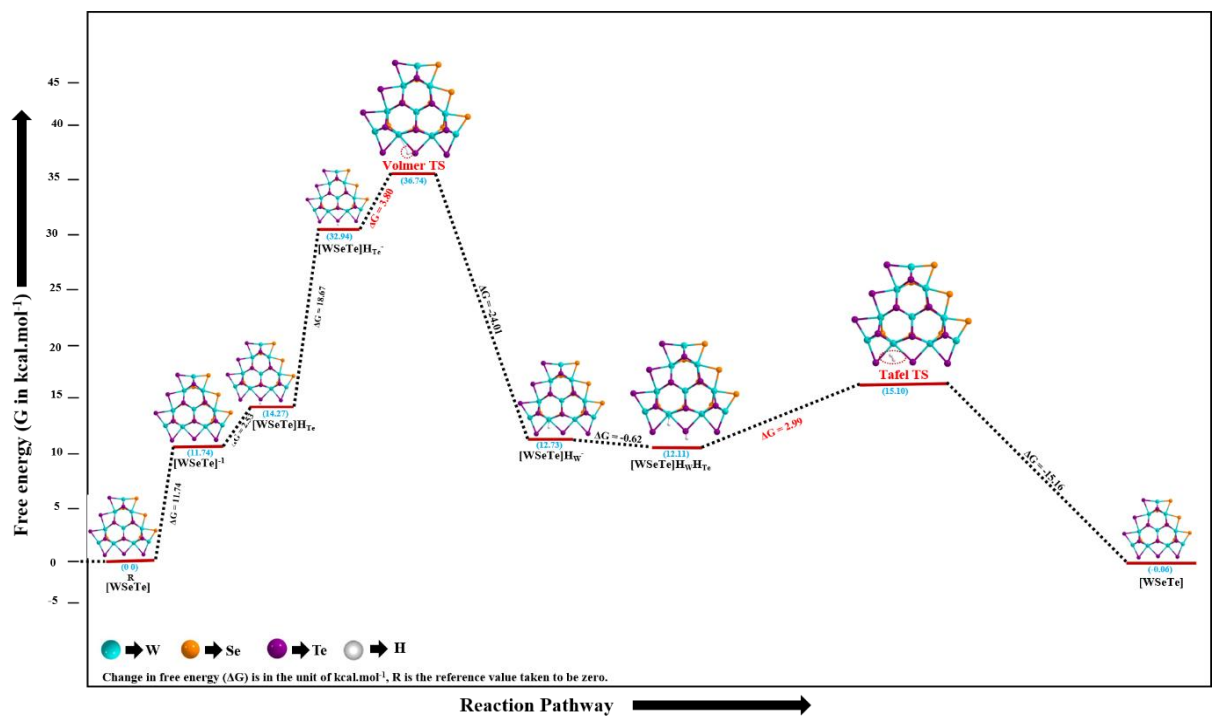

**Figure 8.** The PES of the Volmer-Tafel reaction mechanism during HER on the surface of 2D Janus WSeTe material is depicted here. Changes in free energy (**ΔG**) plots are expressed in kcal.mol$^{-1}$.

It is worth noting here that the TS3 of the Tafel reaction phase in the Volmer-Tafel reaction mechanism is lower than the Heyrovsky transition state of the Volmer-Heyrovsky reaction mechanism in both the gaseous and solvent phases. Here, the present study shows that the Volmer-Heyrovsky mechanism is energetically less favorable than the Volmer-Tafel mechanism for the HER on the surface of the 2D monolayer WSeTe Janus TMD electrocatalyst. This lower reaction barrier proposes that the evolution of H$_2$ can proceed through this Volmer-Tafel pathway with a reaction barrier comparable to that of noble metal-based electrocatalysts.

**OTHER THERMODYNAMIC PROPERTIES OF HER MECHANISM**

**HOMO and LUMO Calculations**

The visualization of the highest occupied molecular orbital (HOMO) and the lowest unoccupied molecular orbital (LUMO) provides a deeper understanding of the $H^*$-migration during the Volmer reaction step, and the electron role in $H_2$ formation and evolution during Heyrovsky or Tafel reaction steps calculations of their respective TS structures, as shown in Figure 9. Here, the energies of the HOMO and the LUMO were calculated at the equilibrium structure of all transition states (TSs) using the same DFT-D method. Figures 9a-b represents the $H^*$-migrated HOMO and LUMO of TS1, respectively, where the electron wavefunction density of $H^*$ has shifted from the Te site to the transition metal site (W-site). During the $H^*$ migration, we found the HOMO-LUMO energy gap around 0.145 eV, as shown in Table 5. This HOMO-LUMO gap is a valuable tool for predicting the stability of the electrocatalytic and color of complexes in solution. The HOMO and LUMO of TS2 in the Heyrovsky reaction step of $H_2$ evolution are represented in Figures 9c-d, respectively. During TS2, the formation of $H_2$ is stabilized due to better overlap between the 5d orbital of the W atom and the 1s orbital of the $H_2$ molecule. The HOMO-LUMO energy gap in the TS2 has been found to be 0.567 eV, as shown in Table 5. The HOMO and LUMO of TS3 in the Tafel reaction step of $H_2$ evolution are represented in Figures 9-f, respectively. The HOMO-LUMO energy gap in the TS3 has been found to be 0.642 eV, as shown in Table 5. The orbital overlap of molecular orbitals during the $H^*$-migration in the Volmer reaction step and the formation of $H_2$ in the Heyrovsky and Tafel reaction steps also revealed the excellent electrocatalytic activity of the 2D monolayer Janus WSeTe for HER.

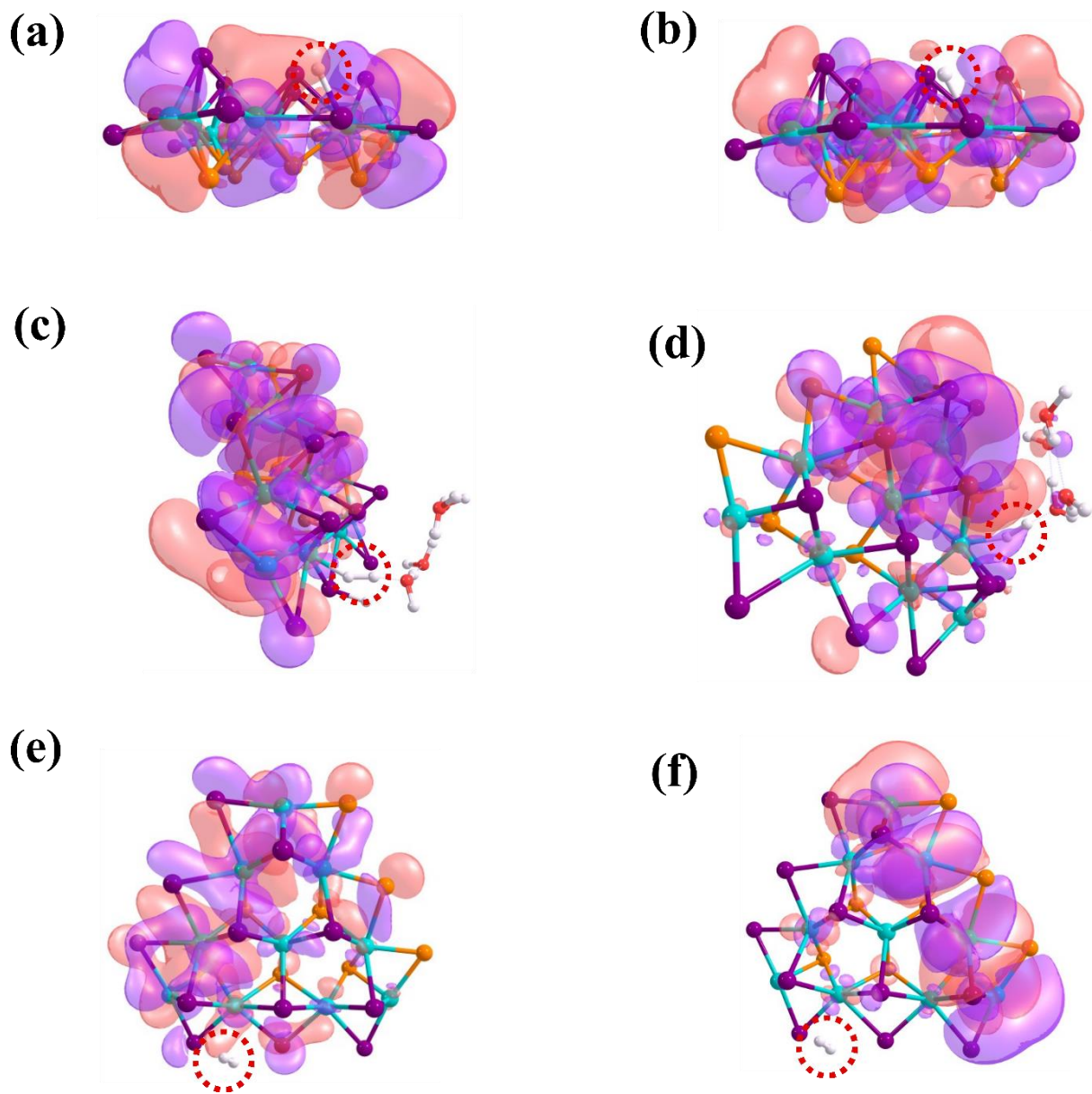

**Figure 9.** (a) HOMO of TS1 for the H*-migration, (b) LUMO of TS1 for the H*-migration, (c) HOMO of TS2 during the Heyrovsky reaction step for $H_2$ formation, (d) LUMO of TS2 during the Heyrovsky reaction step for $H_2$ formation, (e) HOMO of TS3 during the Tafel reaction step for $H_2$ formation and (f) LUMO of TS3 during the Tafel reaction step for $H_2$ formation for 2D Janus WSeTe material are shown here.

**Table 5.** HOMO and LUMO energy and HOMO-LUMO energy gap ($E_g$) of all Transition states (TSs).

| Activation energy barrier | HOMO energy (in eV) | LUMO energy (in eV) | HOMO-LUMO energy gap ($E_g$) (in eV) |
|---|---|---|---|
| H*-migration TS1 | -2.195 | -2.050 | 0.145 |
| Heyrovsky TS3 | -7.274 | -6.707 | 0.567 |
| Tafel TS3 | -5.154 | -4.512 | 0.642 |

**Turnover frequency (TOF) and Tafel Slope Calculation**

To understand the activity of an electrocatalyst surface, turnover frequency (TOF) is the most important factor which influences the overall activity of the HER. The better catalytic activity of an electrocatalyst requires a high turnover frequency, so one must focus on catalysts with higher TOF values. With transition state theory (TST)[70], the TOF at a specific temperature is theoretically given by;

$$TOF = \left(\frac{k_B T}{h}\right) e^{\left(\frac{-\Delta G}{RT}\right)}$$

$k_B = Boltzmann\ constant (3.298 \times 10^{-27}\ kcal.mol^{-1})$

$T = absolute\ temprature\ (here\ 298.15\ K)$

$h = planck's\ constant\ (1.584 \times 10^{-37}\ kcal.sec)$

$R = Universal\ gas\ constant\ (1.987 \times 10^{-3}\ kcal.K^{-1}.mol^{-1})$

$\Delta G = free\ energy\ barrier$

Our study on 2D monolayer Janus WSeTe TMD shows that the TOF at the W-edge is about 2.51 x 10$^6$ sec$^{-1}$ in the solvent phase calculation of the Volmer-Heyrovsky reaction mechanism, corresponding to an activation free energy barrier of 8.72 kcal.mol$^{-1}$. Similarly, the calculated TOF for the solvent phase in the Volmer-Tafel reaction mechanism was found to be 2.48 x 10$^{10}$ sec$^{-1}$, corresponding to an activation energy barrier of 3.27 kcal.mol$^{-1}$ during the Tafel reaction step. Such a large TOF value indicates the excellent performance of the 2D monolayer Janus WSeTe TMD for efficient HER.

Another important parameter for the HER calculation to compare with the experimental results is the Tafel slope. The Tafel slope is the inverse measure of the response strength of the reaction rate to a change in potential.[71] Tafel slope was calculated by the following formula:

$$m = 2.303 \left(\frac{RT}{nF}\right)$$

$$R = \text{Universal gas constant} \ (1.987 \times 10^{-3} \ kcal.K^{-1}.mol^{-1})$$
$$T = \text{absolute temprature} \ (\text{here} \ 298.15 \ K)$$
$$F = \text{faraday constant} \ (96485 \ C.mol^{-1})$$
$$n = \text{number of electrons transferred to the system}$$

The Tafel slope provides information about the catalyst's rate-determining steps, kinetics, and the energy required to obtain the desired activity. For our aperiodic cluster model calculations, the Tafel slope was found to be 29.57 mV.dec$^{-1}$, as n=2 of two electrons were transferred into the system to evolve an H$_2$ molecule.

**Table 6.** Reaction barriers in both gas and solvent phases for various 2D TMDs.

| Catalysts | H* migration TS1 barrier | | Heyrovsky TS2 barrier | | Tafel TS3 barrier | | References |
|---|---|---|---|---|---|---|---|
| | Gas phase (kcal.mol$^{-1}$) | Solvent phase (kcal.mol$^{-1}$) | Gas phase (kcal.mol$^{-1}$) | Solvent phase (kcal.mol$^{-1}$) | Gas phase (kcal.mol$^{-1}$) | Solvent phase (kcal.mol$^{-1}$) | |
| Mn-MoS$_2$ | 7.23 | 10.34 | 10.59 | 10.79 | 90.13 | 93.72 | 2 |
| MoSSe | 3.93 | 7.10 | 5.61 | 4.72 | 8.52 | - | 1 |
| WSeTe | 3.80 | 3.65 | 5.95 | 8.72 | 2.99 | 3.27 | Present work |

The present DFT calculations show that the 2D monolayer Janus WSeTe material has a lower reaction energy barrier for H*-migration in the Volmer reaction step and H$_2$ formation in the Heyrovsky and Tafel reaction step for the HER mechanism compared to other materials listed in Table 6.[1,2] The current DFT studies show that the activation energy barriers for the Volmer reaction during H*-migration across the surface of 2D Janus WSeTe material were around 3.80 kcal.mol$^{-1}$ in the gas phase and 3.65 kcal.mol$^{-1}$ in the solvent phase calculations, respectively. In the same way, considering water as the solvent, the activation energy barriers for the formation of H$_2$ in the Heyrovsky reaction step are approximately 5.95 kcal.mol$^{-1}$ in the gas phase and 8.72 kcal.mol$^{-1}$ in the solvent phase, respectively. Furthermore, the activation energy barriers for forming H$_2$ in the Tafel reaction step are approximately 2.99 kcal.mol$^{-1}$ in the gas phase and 3.27 kcal.mol$^{-1}$ in the solvent phase, respectively. These lowest activation energy barriers and high turnover frequency (TOF) during H* migration and H$_2$ formation confirm the superior HER catalytic activity of 2D monolayer Janus WSeTe compared to other 2D TMDs, as shown in Table 6.

**CONCLUSION**

In conclusion, we have computationally designed a Janus 2D monolayer slab of WSeTe and investigated its electronic and structural properties using the first principle-based DFT-D3 (here, B3LYP-D3 method) calculations. We discovered Janus WSeTe as an efficient electrocatalyst for the HER mechanism. Compared to pristine $WSe_2$, the Janus structure introduces asymmetry, which helps in the transfer of electrons readily during the reactions. A direct band gap has been found to be 2.39 eV for the 2D Janus WSeTe monolayer. The lower bandgap of 2D monolayer Janus WSeTe suggests that it can be used as an electrocatalyst for HER. The catalytic performance of 2D monolayer WSeTe was studied based on the change in HER intermediates' adsorption energy in the non-periodic cluster model Janus WSeTe ($W_{10}Se_9Te_{12}$) surface. Each step involved in Volmer-Heyrovsky and Volmer-Tafel mechanism has been performed to study the HER pathway by the M06-L DFT method. The Volmer reaction during $H^*$-migration from Te-site to W-site across the 2D Janus WSeTe material surface was around 3.80 kcal.mol$^{-1}$ in the gas phase and 3.65 kcal.mol$^{-1}$ in the solvent phase calculations, respectively. The activation energy barriers for the formation of $H_2$ in the Heyrovsky reaction step are approximately 5.95 kcal.mol$^{-1}$ in the gas phase and 8.72 kcal.mol$^{-1}$ in the solvent phase, respectively. Furthermore, in the Volmer-Tafel HER mechanism, the activation energy barriers for the formation of $H_2$ in the Tafel reaction step are approximately 2.99 kcal.mol$^{-1}$ in the gas phase and 3.27 kcal.mol$^{-1}$ in the solvent phase, respectively. Here, it concludes that the Volmer-Heyrovsky mechanism is less favorable than the Volmer-Tafel mechanism (due to the lower energy barrier of the Tafel TS3 compared to the Heyrovsky TS3) on the surface of 2D monolayer Janus WSeTe for the HER.

In HOMO and LUMO calculations, TS shows $H^*$-migration in Volmer steps and $H_2$ evolution in Heyrovsky steps or Tafel steps. Again, this calculation explains why the 2D monolayer Janus WSeTe material has excellent HER catalytic activity. The TOF is about 2.51 x 10$^6$ sec$^{-1}$ through the solvent phase calculation of the Volmer-Heyrovsky reaction mechanism, and in the Volmer-Tafel reaction mechanism, TOF was found to be 2.48 x 10$^{10}$ sec$^{-1}$ in the solvent phase. These higher TOF values confirm the effective hydrogen evolution per active site per unit time of the 2D monolayer WSeTe catalyst. The theoretical Tafel slope of our two-electron transfer mechanism is approximately 29.57 mV.dec$^{-1}$. These very low reaction barrier values, low Tafel slope, and very high TOF values confirm the excellent electrocatalytic activity of 2D monolayer Janus WSeTe for HER. It is believed that these results would bring more experimental and theoretical insights to explore the potential of 2D monolayer Janus TMDs-based materials as a catalyst for the HER.

**AUTHORS INFORMATION:**

**Corresponding Author**

**Dr. Srimanta Pakhira -** *Theoretical Condensed Matter Physics and Advanced Computational Materials Science Laboratory, Department of Physics, Indian Institute of Technology Indore (IITI), Khandwa Road, Simrol, Indore, MP 453552, India.*

*Theoretical Condensed Matter Physics and Advanced Computational Materials Science Laboratory, Centre for Advanced Electronics (CAE), Indian Institute of Technology Indore (IITI), Khandwa Road, Simrol, Indore, MP 453552, India.*

ORCID: orcid.org/0000-0002-2488-300X.

Email: spakhira@iiti.ac.in , spakhirafsu@gmail.com

**Authors:**

**Mr. Vikash Kumar -** *Theoretical Condensed Matter Physics and Advanced Computational Materials Science Laboratory, Department of Physics, Indian Institute of Technology Indore (IITI), Khandwa Road, Simrol, Indore, MP 453552, India.*

ORCID: orcid.org/0000-0002-8811-0583

**Mr. Shrish Nath Upadhyay** − *Theoretical Condensed Matter Physics and Advanced Computational Materials Science Laboratory, Discipline of Metallurgical Engineering and Materials Science (MEMS), Indian Institute of Technology Indore (IITI), Khandwa Road, Simrol, Indore, MP 453552, India.*
ORCID: orcid.org/0000-0003-0029-4160.

**Mr. Dikeshwar Halba** − *Theoretical Condensed Matter Physics and Advanced Computational Materials Science Laboratory, Department of Physics, Indian Institute of Technology Indore (IIT Indore), Simrol, Khandwa Road, Indore, Madhya Pradesh 453552, India.*
ORCID: orcid.org/0000-0003-1493-5934.

† **Electronic supplementary information (ESI) available:**

Optimized .cif file of the 2D Janus WSeTe Transition Metal Dichalcogenides has been given in the Supplementary Information in section S1. Detailed information about the relative Gibbs free energy ($\Delta G$), enthalpy ($\Delta H$), and electronic energy ($\Delta E$) calculations for all the reactants, products, intermediates, and TSs during the HER process has been given in Supplementary Information in section S2.

## Author Contributions:

Dr. Pakhira developed the complete idea of this current research work, and Dr. Pakhira, Mr. Vikash Kumar, and Mr. Shrish Nath Upadhyay computationally studied the electronic structures and properties of the 2D WSeTe JTMD. Dr. Pakhira, Mr. Vikash Kumar, and Mr. Shrish Nath Upadhyay explored the whole reaction pathways; intermediates, transition states, and reaction barriers, and Dr. Pakhira explained the HER mechanism by the DFT calculations. Quantum calculations and theoretical models were designed and performed by Dr. Pakhira, Mr. Vikash Kumar, and Mr. Shrish Nath Upadhyay. Dr. Pakhira, Mr. Vikash Kumar, and Mr. Shrish Nath Upadhyay wrote the whole manuscript and prepared all the tables and figures in the manuscript. Mr. Halba helped Dr. Pakhira, Mr. Vikash Kumar, and Mr. Shrish Nath Upadhyay to organize the manuscript.

**Notes**

The authors announce that there are no competing financial interests.

**ACKNOWLEDGEMENT**

The authors are very much grateful to the Science and Engineering Research Board Department of Science and Technology (SERB-DST), Government of India, under Grant No. ECR/2018/000255 and CRG/2021/000572 for providing financial support for this research work. Dr. Srimanta Pakhira acknowledges the SERB-DST for his Early Career Research Award (ECRA) under project number ECR/2018/000255, and for his highly prestigious Ramanujan Faculty Fellowship under scheme number SB/S2/RJN-067/2017. Dr. Pakhira thanks the SERB for providing a highly prestigious Core Research Grant (CRG), SERB-DST, Govt. of India under the scheme number CRG/2021/000572. Mr. Vikash Kumar thanks the Indian Institute of Technology Indore (IIT Indore) and UGC, Govt. of India, for availing his doctoral fellowship UGC Ref. No: 1403/ (CSIR-UGC NET JUNE 2019).  Mr. Upadhyay thanks the Indian Institute of Technology Indore, MHRD, Govt. of India, for providing the doctoral fellowship. Mr. Halba thanks the CSIR, Govt. of India, Govt. of India for providing his doctoral fellowship under scheme no. CSIRAWARD/JRF-NET2022/1208. The author would like to acknowledge the SERB-DST for providing the computing cluster and programs and IIT Indore for providing the basic infrastructure to conduct this research work. We acknowledge the National Supercomputing Mission (NSM) for providing computing resources of 'PARAM Brahma' at IISER Pune, which is implemented by C-DAC and supported by the Ministry of Electronics and Information Technology (MeitY) and Department of Science and Technology (DST), Government of India.

## Conflicts of Interest:

The authors have no additional conflicts of interest.

## References:

(1)     Pakhira, S.; Upadhyay, S. N. Efficient Electrocatalytic H2 Evolution Mediated by 2D Janus MoSSe Transition Metal Dichalcogenide. *Sustain. Energy & Fuels* **2022**, *6*, 1333-1752.

(2)     Ekka, J.; Upadhyay, S. N.; Keil, F. J.; Pakhira, S.  Unveiling the Role of 2D Monolayer Mn-Doped $MoS_2$ Material: Toward an Efficient Electrocatalyst for $H_2$ Evolution Reaction . *Phys. Chem. Chem. Phys.* **2022**, *24*, 265-280.

(3)     Ball, M.; Weeda, M. The Hydrogen Economy-Vision or Reality? *Int. J. Hydrogen Energy* **2015**, *40*, 7903–7919.

(4)     Yang, J.; Shin, H. S. Recent Advances in Layered Transition Metal Dichalcogenides for Hydrogen Evolution Reaction. *J. Mater. Chem. A* **2014**, *2*, 5979–5985.

(5)     Merki, D.; Hu, X. Recent Developments of Molybdenum and Tungsten Sulfides as

Hydrogen Evolution Catalysts. *Energy & Environ. Sci.* **2011**, *4*, 3878–3888.

(6) Lei, Y.; Pakhira, S.; Fujisawa, K.; Wang, X.; Iyiola, O. O.; Perea López, N.; Laura Elías, A.; Pulickal Rajukumar, L.; Zhou, C.; Kabius, B.; Alem, N.; Endo, M.; Lv, R.; Mendoza-Cortes, J. L.; Terrones, M. Low-Temperature Synthesis of Heterostructures of Transition Metal Dichalcogenide Alloys ($W_xMo_{1-x}S_2$) and Graphene with Superior Catalytic Performance for Hydrogen Evolution. *ACS Nano* **2017**, *11*, 5103–5112.

(7) Upadhyay, S. N.; Satrughna, J. A. K.; Pakhira, S. Recent Advancements of Two-Dimensional Transition Metal Dichalcogenides and Their Applications in Electrocatalysis and Energy Storage. *Emergent Mater.* **2021**, *4*, 951–970.

(8) Chia, X.; Sutrisnoh, N. A. A.; Sofer, Z.; Luxa, J.; Pumera, M. Morphological Effects and Stabilization of the Metallic 1T Phase in Layered V-, Nb-, and Ta-Doped $WSe_2$ for Electrocatalysis. *Chem. Eur. J.* **2018**, *24*, 3199–3208.

(9) Ju, L.; Bie, M.; Tang, X.; Shang, J.; Kou, L. Janus WSSe Monolayer: An Excellent Photocatalyst for Overall Water Splitting. *ACS Appl. Mater. & Interfaces* **2020**, *12*, 29335–29343.

(10) Dong, L.; Lou, J.; Shenoy, V. B. Large In-Plane and Vertical Piezoelectricity in Janus Transition Metal Dichalchogenides. *ACS Nano* **2017**, *11*, 8242–8248.

(11) Li, R.; Cheng, Y.; Huang, W. Recent Progress of Janus 2D Transition Metal Chalcogenides: From Theory to Experiments. *Small* **2018**, *14*, 1802091.

(12) Som, N. N.; Jha, P. K. Hydrogen Evolution Reaction of Metal Di-Chalcogenides: $ZrS_2$, $ZrSe_2$ and Janus ZrSSe. *Int. J. Hydrogen Energy* **2020**, *45*, 23920–23927.

(13) Er, D.; Ye, H.; Frey, N. C.; Kumar, H.; Lou, J.; Shenoy, V. B. Prediction of Enhanced Catalytic Activity for Hydrogen Evolution Reaction in Janus Transition Metal Dichalcogenides. *Nano Lett.* **2018**, *18*, 3943–3949.

(14) Lu, A.-Y.; Zhu, H.; Xiao, J.; Chuu, C.-P.; Han, Y.; Chiu, M.-H.; Cheng, C.-C.; Yang, C.-W.; Wei, K.-H.; Yang, Y.; others. Janus Monolayers of Transition Metal Dichalcogenides. *Nat. Nanotechnol.* **2017**, *12*, 744–749.

(15) Chaurasiya, R.; Gupta, G. K.; Dixit, A. Ultrathin Janus WSSe Buffer Layer for $W(S/Se)_2$ Absorber Based Solar Cells: A Hybrid, DFT and Macroscopic, Simulation Studies. *Sol. Energy Mater. Sol. Cells* **2019**, *201*, 110076.

(16) Maghirang, A. B.; Huang, Z.-Q.; Villaos, R. A. B.; Hsu, C.-H.; Feng, L.-Y.; Florido, E.; Lin, H.; Bansil, A.; Chuang, F.-C. Predicting Two-Dimensional Topological Phases in Janus Materials by Substitutional Doping in Transition Metal Dichalcogenide Monolayers. *npj 2D Mater. Appl.* **2019**, *3*, 1–8.

(17) Cheng, Y. C.; Zhu, Z. Y.; Tahir, M.; Schwingenschlögl, U. Spin-Orbit--Induced Spin Splittings in Polar Transition Metal Dichalcogenide Monolayers. *EPL (Europhysics Lett.* **2013**, *102*, 57001.

(18) Kumar, V.; Pakhira, S. Mechanistic Understanding of Efficient Electrocatalytic Hydrogen Evolution Reaction on 2D Monolayer WSSe Janus Transition Metal Dichalcogenide. *Mol. Syst. Des. Eng.* **2023**.

(19) Becke, A. D. Density-Functional Thermochemistry. III. The Role of Exact Exchange. *J. Chem. Phys.* **1993**, *98*, 5648–5652.

(20) Pakhira, S.; Mendoza-Cortes, J. L. Tuning the Dirac Cone of Bilayer and Bulk Structure Graphene by Intercalating First Row Transition Metals Using First-Principles Calculations. *J. Phys. Chem. C* **2018**, *122*, 4768–4782.

(21) Li, F.; Wei, W.; Zhao, P.; Huang, B.; Dai, Y. Electronic and Optical Properties of Pristine and Vertical and Lateral Heterostructures of Janus MoSSe and WSSe. *J. Phys. Chem. Lett.* **2017**, *8*, 5959–5965.

(22) Niu, W.; Pakhira, S.; Marcus, K.; Li, Z.; Mendoza-Cortes, J. L.; Yang, Y. Apically Dominant Mechanism for Improving Catalytic Activities of N-Doped Carbon Nanotube Arrays in Rechargeable Zinc–Air Battery. *Adv. Energy Mater.* **2018**, *8*, 18000480.

(23) Pakhira, S.; Sen, K.; Sahu, C.; Das, A. K. Performance of Dispersion-Corrected Double Hybrid Density Functional Theory: A Computational Study of OCS-Hydrocarbon van Der Waals Complexes. *J. Chem. Phys.* **2013**, *138*, 164319.

(24) Sinha, N.; Pakhira, S. Tunability of the Electronic Properties of Covalent Organic Frameworks. *ACS Appl. Electron. Mater.* **2021**, *3*, 720–732.

(25) Becke, A. D. Density - Functional Thermochemistry . III . The Role of Exact Exchange Density-Functional Thermochemistry . III . The Role of Exact Exchange. *J. Chern. Phys.* **2005**, *98*, 5648–5652.

(26) Upadhyay, S. N.; Pakhira, S. Mechanism of Electrochemical Oxygen Reduction

Reaction at Two-Dimensional Pt-Doped MoSe$_2$ Material: An Efficient Electrocatalyst. *J. Mater. Chem. C* **2021**, *9*, 11331–11342.

(27) Puttaswamy, R.; Nagaraj, R.; Kulkarni, P.; Beere, H. K.; Upadhyay, S. N.; Balakrishna, R. G.; Sanna Kotrappanavar, N.; Pakhira, S.; Ghosh, D. Constructing a High-Performance Aqueous Rechargeable Zinc-Ion Battery Cathode with Self-Assembled Mat-like Packing of Intertwined Ag (I) Pre-Inserted V$_3$O$_7$·H$_2$O Microbelts with Reduced Graphene Oxide Core. *ACS Sustain. Chem. & Eng.* **2021**, *9*, 3985–3995.

(28) Dovesi, R.; Erba, A.; Orlando, R.; Zicovich-Wilson, C. M.; Civalleri, B.; Maschio, L.; Rérat, M.; Casassa, S.; Baima, J.; Salustro, S.; others. Quantum-Mechanical Condensed Matter Simulations with CRYSTAL. *Wiley Interdiscip. Rev. Comput. Mol. Sci.* **2018**, *8*, e1360.

(29) Patel, C.; Singh, R.; Dubey, M.; Pandey, S. K.; Upadhyay, S. N.; Kumar, V.; Sriram, S.; Than Htay, M.; Pakhira, S.; Atuchin, V. V; Mukherjee, S. Large and Uniform Single Crystals of MoS$_2$ Monolayers for Ppb-Level NO$_2$ Sensing. *ACS Appl. Nano Mater.* **2022**, *7*, 9415-9426.

(30) Upadhyay, S. N.; Sardar, V. B.; Singh, A.; Kumar, V.; Pakhira, S. Elucidating the Oxygen Reduction Reaction Mechanism on the Surfaces of 2D Monolayer CsPbBr$_3$ Perovskite. *Phys. Chem. Chem. Phys.* **2022**, *24*, 28283–28294.

(31) Nagaraj, R.; Pakhira, S.; Aruchamy, K.; Yadav, P.; Mondal, D.; Dharmalingm, K.; Sanna Kotrappanavar, N.; Ghosh, D. Catalyzing the Intercalation Storage Capacity of Aqueous Zinc-Ion Battery Constructed with Zn (II) Preinserted Organo-Vanadyl Hybrid Cathode. *ACS Appl. Energy Mater.* **2020**, *3*, 3425–3434.

(32) Pakhira, S.; Lucht, K. P.; Mendoza-Cortes, J. L. Iron Intercalation in Covalent-Organic Frameworks: A Promising Approach for Semiconductors. *J. Phys. Chem. C* **2017**, *121*, 21160–21170.

(33) Hui, J.; Pakhira, S.; Bhargava, R.; Barton, Z. J.; Zhou, X.; Chinderle, A. J.; Mendoza-Cortes, J. L.; Rodríguez-López, J. Modulating Electrocatalysis on Graphene Heterostructures: Physically Impermeable Yet Electronically Transparent Electrodes. *ACS Nano* **2018**, *12*, 2980–2990.

(34) Liang, K.; Pakhira, S.; Yang, Z.; Nijamudheen, A.; Ju, L.; Wang, M.; Aguirre-Velez, C.

I.; Sterbinsky, G. E.; Du, Y.; Feng, Z.; others. S-Doped MoP Nanoporous Layer toward High-Efficiency Hydrogen Evolution in PH-Universal Electrolyte. *ACS Catal.* **2018**, *9*, 651–659.

(35) Grimme, S.; Antony, J.; Ehrlich, S.; Krieg, H. A Consistent and Accurate Ab Initio Parametrization of Density Functional Dispersion Correction (DFT-D) for the 94 Elements H-Pu. *J. Chem. Phys.* **2010**, *132*, 154104.

(36) Pakhira, S.; Takayanagi, M.; Nagaoka, M. Diverse Rotational Flexibility of Substituted Dicarboxylate Ligands in Functional Porous Coordination Polymers. *J. Phys. Chem. C* **2015**, *119*, 28789–28799.

(37) Caldeweyher, E.; Bannwarth, C.; Grimme, S. Extension of the D3 Dispersion Coefficient Model. *J. Chem. Phys.* **2017**, *147*, 34112.

(38) Lei, Y.; Pakhira, S.; Fujisawa, K.; Liu, H.; Guerrero-Bermea, C.; Zhang, T.; Dasgupta, A.; Martinez, L. M.; Rao Singamaneni, S.; Wang, K.; Shallenberger, J.; Elías, A. L.; Cruz-Silva, R.; Endo, M.; Mendoza-Cortes, J. L.; Terrones, M. Low Temperature Activation of Inert Hexagonal Boron Nitride for Metal Deposition and Single Atom Catalysis. *Mater. Today* **2021**, *51*, 108-116.

(39) Laun, J.; Bredow, T. BSSE-Corrected Consistent Gaussian Basis Sets of Triple-Zeta Valence with Polarization Quality of the Sixth Period for Solid-State Calculations. *J. Comput. Chem.* **2021**, *42*, 1064–1072.

(40) Vilela Oliveira, D.; Laun, J.; Peintinger, M. F.; Bredow, T. BSSE-Correction Scheme for Consistent Gaussian Basis Sets of Double-and Triple-Zeta Valence with Polarization Quality for Solid-State Calculations. *J. Comput. Chem.* **2019**, *40*, 2364–2376.

(41) Peintinger, M. F.; Oliveira, D. V.; Bredow, T. Consistent Gaussian Basis Sets of Triple-Zeta Valence with Polarization Quality for Solid-State Calculations. *J. Comput. Chem.* **2013**, *34*, 451–459.

(42) Pakhira, S.; Takayanagi, M.; Nagaoka, M. Diverse Rotational Flexibility of Substituted Dicarboxylate Ligands in Functional Porous Coordination Polymers. *J. Phys. Chem. C* **2015**, *119*, 28789–28799.

(43) Pakhira, S.; Mendoza-Cortes, J. L. Quantum Nature in the Interaction of Molecular Hydrogen with Porous Materials: Implications for Practical Hydrogen Storage. *J. Phys.*

*Chem. C* **2020**, *124*, 6454–6460.

(44) Montoya, A.; Truong, T. N.; Sarofim, A. F. Spin Contamination in Hartree-Fock and Density Functional Theory Wavefunctions in Modeling of Adsorption on Graphite. *J. Phys. Chem. A* **2000**, *104*, 6108–6110.

(45) Baker, J.; Scheiner, A.; Andzelm, J. Spin Contamination in Density Functional Theory. *Chem. Phys. Lett.* **1993**, *216*, 380–388.

(46) Salustro, S.; Ferrari, A. M.; Orlando, R.; Dovesi, R. Comparison between Cluster and Supercell Approaches: The Case of Defects in Diamond. *Theor. Chem. Acc.* **2017**, *136*, 1–13.

(47) Evarestov, R. A.; Smirnov, V. P. Modification of the Monkhorst-Pack Special Points Meshes in the Brillouin Zone for Density Functional Theory and Hartree-Fock Calculations. *Phys. Rev. B* **2004**, *70*, 233101.

(48) Momma, K.; Izumi, F. VESTA 3 for Three-Dimensional Visualization of Crystal, Volumetric and Morphology Data. *J. Appl. Crystallogr.* **2011**, *44*, 1272–1276.

(49) Zhao, Y.; Truhlar, D. G. A New Local Density Functional for Main-Group Thermochemistry, Transition Metal Bonding, Thermochemical Kinetics, and Noncovalent Interactions. *J. Chem. Phys.* **2006**, *125*, 194101.

(50) Zhao, Y.; Truhlar, D. G. The M06 Suite of Density Functionals for Main Group Thermochemistry, Thermochemical Kinetics, Noncovalent Interactions, Excited States, and Transition Elements: Two New Functionals and Systematic Testing of Four M06-Class Functionals and 12 Other Function. *Theor. Chem. Acc.* **2008**, *120*, 215–241.

(51) Frisch, M. J.; Trucks, G. W.; Schlegel, H. B.; Scuseria, G. E.; Robb, M. A.; Cheeseman, J. R.; Scalmani, G.; Barone, V.; Petersson, G. A.; Nakatsuji, H. . Gaussian 16, Revision C. 01; GaussianNo Title. *Inc. Wallingford CT* **2016**.

(52) Sinha, N.; Pakhira, S. $H_2$ Physisorption on Covalent Organic Framework Linkers and Metalated Linkers: A Strategy to Enhance Binding Strength. *Mol. Syst. Des. & Eng.* **2022**, *7*, 577–591.

(53) Hay, P. J.; Wadt, W. R. Ab Initio Effective Core Potentials for Molecular Calculations. Potentials for K to Au Including the Outermost Core Orbitale. *J. Chem. Phys.* **1985**, *82*, 299–310.

(54) Hay, P. J.; Wadt, W. R. Ab Initio Effective Core Potentials for Molecular Calculations. Potentials for the Transition Metal Atoms Sc to Hg. *J. Chem. Phys.* **1985**, *82*, 270–283.

(55) Frisch, M. J.; Pople, J. A.; Binkley, J. S. Self-Consistent Molecular Orbital Methods 25. Supplementary Functions for Gaussian Basis Sets. *J. Chem. Phys.* **1984**, *80*, 3265–3269.

(56) Hehre, W. J.; Ditchfield, R.; Pople, J. A. Self—Consistent Molecular Orbital Methods. XII. Further Extensions of Gaussian—Type Basis Sets for Use in Molecular Orbital Studies of Organic Molecules. *J. Chem. Phys.* **1972**, *56*, 2257–2261.

(57) Pakhira, S.; Singh, R. I.; Olatunji-Ojo, O.; Frenklach, M.; Lester, W. A. Quantum Monte Carlo Study of the Reactions of CH with Acrolein: Major and Minor Channels. *J. Phys. Chem. A* **2016**, *120*, 3602–3612.

(58) Pakhira, S.; Kumar, V.; Ghosh, S. Revealing the Superior Electrocatalytic Performance of 2D Monolayer $WSe_2$ Transition Metal Dichalcogenide for Efficient $H_2$ Evolution Reaction. *Adv. Mater. Interfaces* **2023**, *10*, 2202075.

(59) Upadhyay, S. N.; Pakhira, S. Nanostructured Pt-Doped 2D $MoSe_2$: An Efficient Bifunctional Electrocatalyst for Both Hydrogen Evolution and Oxygen Reduction Reactions. *Phys. Chem. Chem. Phys.* **2022**, *24*, 22823–22844.

(60) Andrienko, G. A. Chemcraft – graphical software for visualization of quantum chemistry computations. *www http//www. chemcraftprog. com* **2015**.

(61) Uematsu, M.; Frank, E. U. Static Dielectric Constant of Water and Steam. *J.Phys.Chem. Ref. Data* **1980**, *9*, 1291–1306.

(62) Garza, A. J.; Pakhira, S.; Bell, A. T.; Mendoza-Cortes, J. L.; Head-Gordon, M. Reaction Mechanism of the Selective Reduction of $CO_2$ to CO by a Tetraaza $[Co^{II}N_4H]^{2+}$ Complex in the Presence of Protons. *Phys. Chem. Chem. Phys.* **2018**, *20*, 24058–24064.

(63) Vu, T. V; Hieu, N. V; Phuc, H. V; Hieu, N. N.; Bui, H. D.; Idrees, M.; Amin, B.; Nguyen, C. V. Graphene/WSeTe van Der Waals Heterostructure: Controllable Electronic Properties and Schottky Barrier via Interlayer Coupling and Electric Field. *Appl. Surf. Sci.* **2020**, *507*, 145036.

(64) Xia, C.; Xiong, W.; Du, J.; Wang, T.; Peng, Y.; Li, J. Universality of Electronic Characteristics and Photocatalyst Applications in the Two-Dimensional Janus Transition Metal Dichalcogenides. *Phys. Rev. B* **2018**, *98*, 1–8.

(65) Shi, W.; Wang, Z. Mechanical and Electronic Properties of Janus Monolayer Transition Metal Dichalcogenides. *J. Phys. Condens. Matter* **2018**, *30*, 215301.

(66) Yang, Y.; Zhang, Y.; Ye, H.; Yu, Z.; Liu, Y.; Su, B.; Xu, W. Structural and Electronic Properties of 2H Phase Janus Transition Metal Dichalcogenide Bilayers. *Superlattices Microstruct.* **2019**, *131*, 8–14.

(67) Wang, J.; Liu, J.; Zhang, B.; Ji, X.; Xu, K.; Chen, C.; Miao, L.; Jiang, J. The Mechanism of Hydrogen Adsorption on Transition Metal Dichalcogenides as Hydrogen Evolution Reaction Catalyst. *Phys. Chem. Chem. Phys.* **2017**, *19*, 10125–10132.

(68) Li, C.; Gao, H.; Wan, W.; Mueller, T. Mechanisms for Hydrogen Evolution on Transition Metal Phosphide Catalysts and a Comparison to Pt (111). *Phys. Chem. Chem. Phys.* **2019**, *21*, 24489–24498.

(69) Tissandier, M. D.; Cowen, K. A.; Feng, W. Y.; Gundlach, E.; Cohen, M. H.; Earhart, A. D.; Coe, J. V.; Tuttle, T. R. The Proton's Absolute Aqueous Enthalpy and Gibbs Free Energy of Solvation from Cluster-Ion Solvation Data. *J. Phys. Chem. A* **1998**, *102*, 7787–7794.

(70) Jordan, P. C. Theories of Reaction Rates. In *Chemical Kinetics and Transport*; Springer, 1979; pp 269–323.

(71) Huang, Y.; Nielsen, R. J.; Goddard III, W. A.; Soriaga, M. P. The Reaction Mechanism with Free Energy Barriers for Electrochemical Dihydrogen Evolution on MoS$_2$. *J. Am. Chem. Soc.* **2015**, *137*, 6692–6698.

**TOC and Graphical Abstract:**

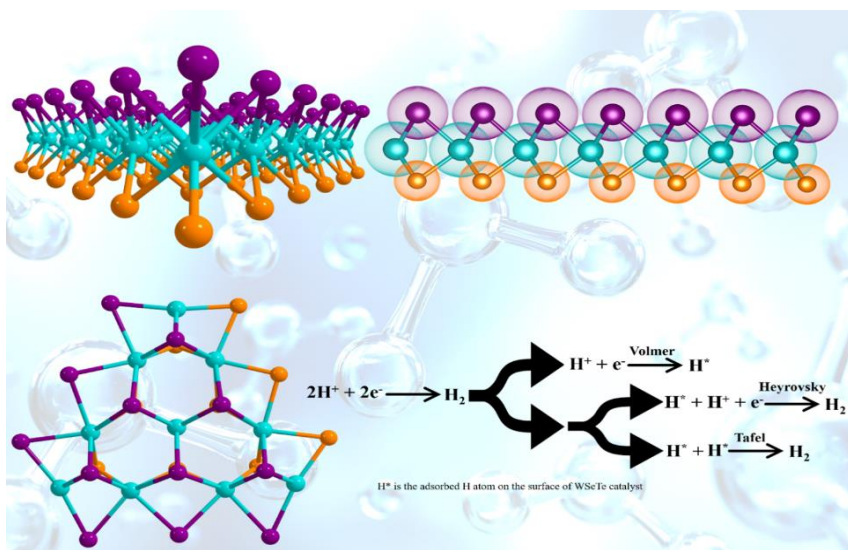